\documentclass[%
 reprint,
superscriptaddress,   
 amsmath,amssymb,
 aps,
 prl
]{revtex4-2}

\usepackage{graphicx}
\usepackage{dcolumn}
\usepackage{bm}

\begin{document}

\preprint{APS/123-QED}

\title{Single-species Atomic Comagnetometer Based on $^{87}\rm Rb$ Atoms}

\author{Zhiguo Wang}
\thanks{These authors contributed equally to this work.}
\affiliation{%
 College of Advanced Interdisciplinary Studies, National University of Defense Technology, Changsha, 410073, P R China
}%
\affiliation{%
 Interdisciplinary Center of Quantum Information, National University of Defense Technology, Changsha 410073, P R China
}
\author{Xiang Peng}%
\thanks{These authors contributed equally to this work.}
\affiliation{
State Key Laboratory of Advanced Optical Communication Systems and Networks, Department of Electronics, and Center for Quantum Information Technology, Peking University, Beijing 100871, P R China
}%
\author{Rui Zhang}
\affiliation{%
 College of Liberal Arts and Sciences, National University of Defense Technology, Changsha, 410073, P R China}%
\author{Hui Luo}
 \email[Corresponding author. Email: ]{luohui.luo@163.com (H.L.)}
\affiliation{%
 College of Advanced Interdisciplinary Studies, National University of Defense Technology, Changsha, 410073, P R China
}%
\affiliation{%
 Interdisciplinary Center of Quantum Information, National University of Defense Technology, Changsha 410073, P R China
}
\author{Jiajia Li}
\affiliation{%
 College of Advanced Interdisciplinary Studies, National University of Defense Technology, Changsha, 410073, P R China
}%
\author{Zhiqiang Xiong}
\affiliation{%
 College of Advanced Interdisciplinary Studies, National University of Defense Technology, Changsha, 410073, P R China
}%
\author{Shanshan Wang}
\affiliation{%
 College of Advanced Interdisciplinary Studies, National University of Defense Technology, Changsha, 410073, P R China
}%
\author{Hong Guo}%
\email[Corresponding author. Email: ]{hongguo@pku.edu.cn (H.G.)}
\affiliation{
State Key Laboratory of Advanced Optical Communication Systems and Networks, Department of Electronics, and Center for Quantum Information Technology, Peking University, Beijing 100871, P R China
}%

\date{\today}

\begin{abstract}
 The comagnetometer has been one of the most sensitive devices with which to test new physics related to spin-dependent interactions, but the comagnetometers based on overlapping ensembles of multiple spin species usually suffer from systematic errors due to magnetic field gradients. Here, we propose a comagnetometer based on the Zeeman transitions of the dual hyperfine levels in ground-state $^{87} \rm Rb$ atoms, which shows nearly negligible sensitivity to variations of laser power and frequency, magnetic field, and magnetic field gradients. We measured the hypothetical spin-dependent gravitational energy of the proton with the comagnetometer, which is smaller than $4\times10^{-18}$ eV, comparable to the most stringent existing constraint.
 Through optimizing the system parameters such as cell temperature, laser power, amplitude of driving magnetic field, as well as choosing better current source, it is possible to improve the sensitivity of the comagnetometer further.
\begin{description}
\item[PACS: 07.55.Ge, 32.60.+I, 32.10.Fn]
\end{description}
\end{abstract}

\maketitle



\section{\label{sec:level1} Introduction}

Atomic comagnetometers based on spin-precession have been widely used for many kinds of fundamental physics experiments \cite{RevModPhys.90.025008}, e.g., probing Lorentz- and CPT- violations \cite{PhysRevLett.112.110801, PhysRevD.60.116010, PhysRevLett.105.151604}, searching for permanent electric dipole moments \cite{YOSHIMI200213, PhysRevLett.72.2363, PhysRevLett.86.22, PhysRevX.7.041034, SATO2018588} and exotic spin-dependent interactions \cite{PhysRevLett.111.102001, PhysRevLett.103.261801, PhysRevLett.111.100801}. Although these comagnetometers use spins of different species overlapping in the same space to suppress the magnetic field variation in common mode \cite{PhysRevLett.72.2363, PhysRevLett.86.22, PhysRevX.7.041034, SATO2018588,YOSHIMI200213, PhysRevLett.112.110801, PhysRevD.60.116010, PhysRevLett.105.151604, PhysRevLett.111.102001, PhysRevLett.103.261801, PhysRevLett.111.100801}, there are still several systematic errors existing in comagnetometers \cite{PhysRevLett.113.163002}.

Nearly all comagnetometers work on gas or liquid atoms, which move randomly in a vessel. The ensemble-averaged positions of different spin species can be spatially separated, due to reasons such as different thermal diffusion rates accompanied with temperature gradient, non-uniform polarization accompanied with different transverse relaxation time, etc. \cite{PhysRevLett.113.163002, Ledbetter_2012}. As a result, the magnetic field fluctuations cannot be common-mode suppressed effectively in the presence of a magnetic field gradient, especially for the comagnetometers with more than one species atoms. Moreover, comagnetometers based on gas spin usually utilize the spin-exchange optical pumping to polarize the spins, which will cause a frequency shift of the Zeeman transitions. For example, in comagnetometers based on alkali-metal and dual noble-gas spins ($\rm Rb-^{129}Xe/^{131}Xe$, $\rm Rb-^{129}Xe/^{3}He$, etc.), polarized alkali-metal atoms exert different effective magnetic fields on the noble-gas spins, which leads to remarkable systematic errors \cite{PhysRevLett.111.102001}. Several new methods have been proposed to overcome this kind of problem. One is to divide the pumping and probing processes into two temporally separated phases and to measure the spin precession of $\rm ^3 He$ and $\rm ^{129} Xe$ in the dark \cite{PhysRevLett.120.033401}. Another is to operate the $\rm ^{129} Xe/^{131}Xe$ spins with synchronous pumping \cite{PhysRevLett.115.253001}. In a comagnetometer based on $\rm ^{85}Rb/ ^{87}Rb$ contained in an evacuated coated cell, the influence of magnetic field gradient can be greatly suppressed due to the fast motion of gas atoms \cite{doi10.1002Kimball, PhysRevD.96.075004, doi10.1002Mora}. With using careful and complicated calibrations or compensations for the effects caused by pumping light, probing light, magnetic field gradients, the systematic errors due to light shift and magnetic field gradients, have been suppressed to some extent \cite{doi10.1002Kimball, PhysRevD.96.075004, doi10.1002Mora}.
To eliminate the systematic errors caused by magnetic field gradients, a comagnetometer with identical molecules has been proposed \cite{PhysRevLett.121.023202}, which reduces the influence of the magnetic field variation and magnetic field gradient efficiently, but the polarization and detection of the nuclear spins in molecules make the system slightly more complex.

In this paper, we propose a single-species comagnetometer based on the hyperfine levels of $\rm^{87} Rb$ in a paraffin-coated cell. It is operated with only a linearly polarized laser beam propagating orthogonal to a bias magnetic field. Owing to the single-species operation, fast motion of atoms as well as nearly identical gyromagnetic ratio of the dual hyperfine levels, the comagnetometer shows an extremely small dependence on the magnetic field gradient. In addition, it exhibits less parameter-sensitive to variations of probe laser power and frequency, since the influence of light shift is small \cite{Zhiguo_2019} \nocite{seltzer2008, mathur1968, chalupczak2010, hu2018, costanzo2016, abrarov2011, Abrarov2020, PhysRev.174.23, budker2007optical, Yang2019, jensen2009cancellation, wang2017comparison, lee2016spin, PhysRevA.37.2877, pendlebury2004geometric}. In our experiment, we obtain remarkable magnetic field variation suppression and less dependence on magnetic field gradient. The preliminary experimental results show that the proposed comagnetometer is a promising device for spin-gravity interaction tests.

\section{\label{sec:level2} Experiments and Results}

The configuration of the comagnetometer is shown in Fig. 1. In this experiment, we use a single-beam double-resonance-alignment-magnetometer configuration \cite{PhysRevA.74.033401, PhysRevApplied.10.034035}. The key component is a spherical glass cell $20\ \rm mm$ in diameter and with a paraffin coating on the internal surface, which was made by Peking University. Benefitting from the paraffin coating, the intrinsic linewidth of the Zeeman spectrum can be as narrow as approximately 2 Hz. A static bias magnetic field $B_0$ along the $\rm z$ axis and an orthogonal driving magnetic field $B_{\rm{x}}$ oscillating at frequency $f$ are applied. A linearly polarized laser beam, more than 1 GHz far detuning from either of the two transitions between $5^{2}S_{1/2} (F = 1,2)$ to  $5^{2}P_{3/2} (F')$, is used to polarize the $^{87}\rm Rb$ atoms and to probe the evolution of the atomic polarization. The magnetic resonance process can be understood in three steps, i.e., preparation, evolution, and probing \cite{Bevilacqua_2016}. At first, the laser beam creates rank two polarizations (alignments) in both the $F=1$ and $F=2$  hyperfine levels of the ground state of the $^{87}\rm Rb$ atom. Because the Larmor frequency at $B_0$ is much larger than the relaxation rate of the atomic polarization, only the alignment component along $B_0$ remains, while other alignment components relax to zero quickly. Then, under the combined actions of the bias field $B_0$, driving field $B_{\rm{x}}$, and relaxation, the atomic alignment precesses around $B_0$ at the driving frequency $f$. When $f$  equals the Larmor frequency of either of the hyperfine levels, the alignment of corresponding hyperfine level comes to magnetic resonance. Finally, the precessing alignments of both hyperfine levels modulate the polarization of the laser beam, which is detected by the Faraday rotation method \cite{Bevilacqua_2016}. All the experiments were done at ambient temperature ($24 \pm 1) ^\circ$C.

\begin{figure}[h]
\includegraphics[width=3.3in]{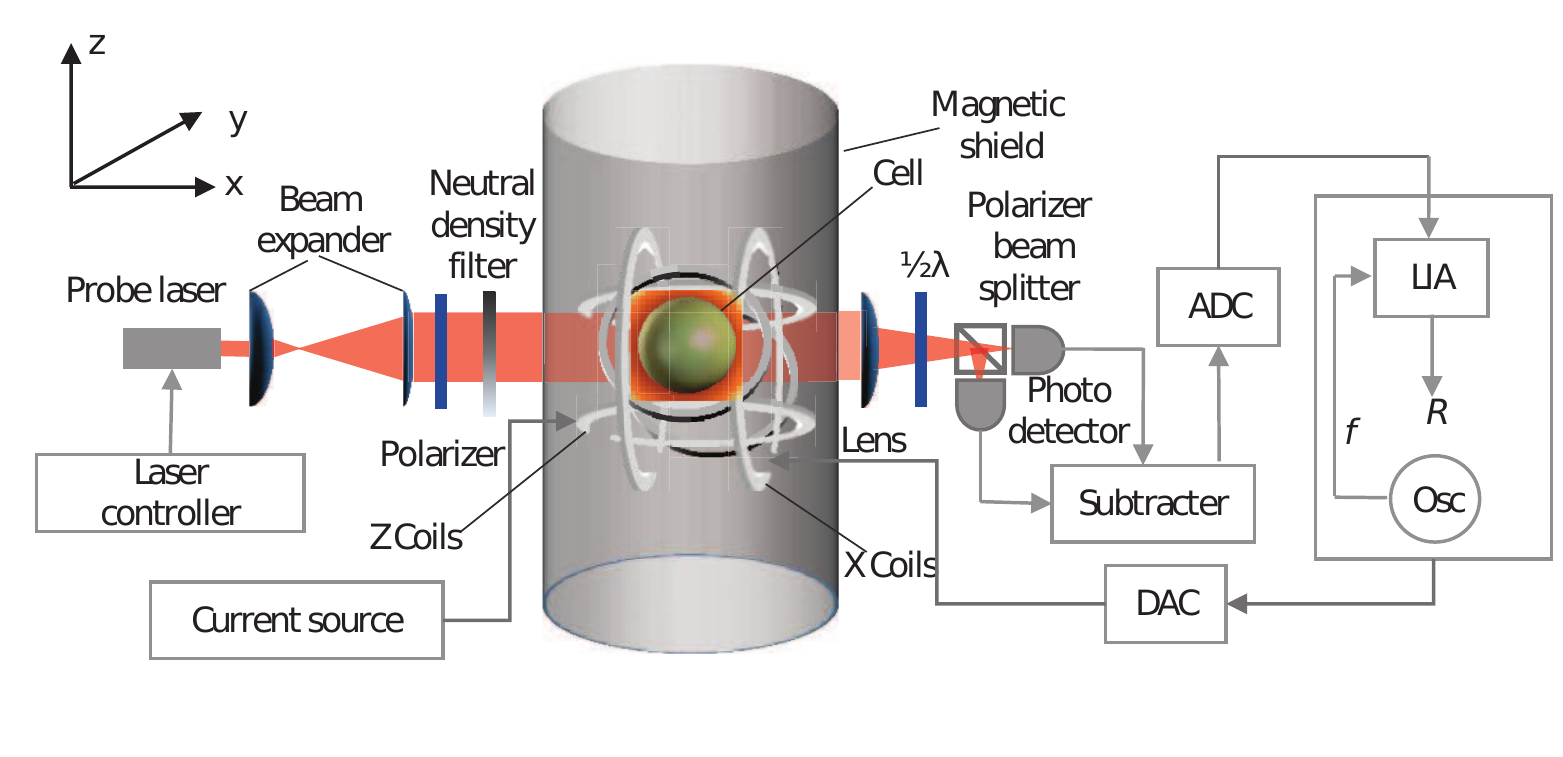}
\caption{\label{fig:setup} (color online). Experimental setup. A spherical glass cell filled with an excess of $^{87}\rm Rb$ is placed in the magnetic field shield. A set of solenoids are used to produce the bias magnetic field in the $\rm z$ direction and the driving magnetic field in the $\rm x$  direction. The probe light from a DFB laser at 780 nm propagates through a linear polarizer (polarization axis along the y axis), a collimator made with two lenses, and a neutral density filter. The probe laser propagates through the cell and then is focused with a lens. A half-wave plate, a polarization beam splitter, and a balanced detector made up of two photodiodes are used to detect the paramagnetic Faraday rotation signal. The difference of the two photodiodes' signal is digitized through an analog-to-digital converter (ADC). The output of an oscillator (OSC) is sent to a digital-to-analog converter (DAC) and then drives the $\rm x$ coils to produce an oscillating magnetic field $B_{\rm{x}}$. We obtain the Zeeman spectrum of the Rb atoms by scanning the frequency of $B_{\rm{x}}$ and recording the R channel of a lock-in amplifier (LIA) made with the $\rm Labview^{TM}$ program.
The amplitude of $B_x$ is set as 0.1 nT and the frequency is changed linearly in the scan process. The $\rm z$ coils are driven by a stable current source, producing $B_0\approx 11.72 \ \rm{\mu T}$. A set of anti-Helmholtz coils is also used to produce the magnetic field gradient $\partial B_{\rm{z}}/\partial \rm{z}$.}
\end{figure}

A typical Zeeman spectrum is shown in Fig. 2(a). The frequency difference for the Zeeman transitions at two hyperfine levels in the ground state is $326.4\ \rm Hz$ with a bias magnetic field of $11.72 \ \rm \mu T$. The ratio $\gamma_2/\gamma_1$ is 0.996033924 in theory \cite{Daniel2015}, where $\gamma_2$ and $\gamma_1$ are the gyromagnetic ratios for $F=2$ and $F=1$, respectively. The resonant frequency for each hyperfine level is obtained by fitting the resonance curve with a Lorentzian profile, as shown in Figs. 2(b) and 2(c). Since the Zeeman transition frequencies of the two hyperfine levels are both proportional to the magnetic field, they can be used as a comagnetometer.

\begin{figure}[h]
\includegraphics[width=3.3in]{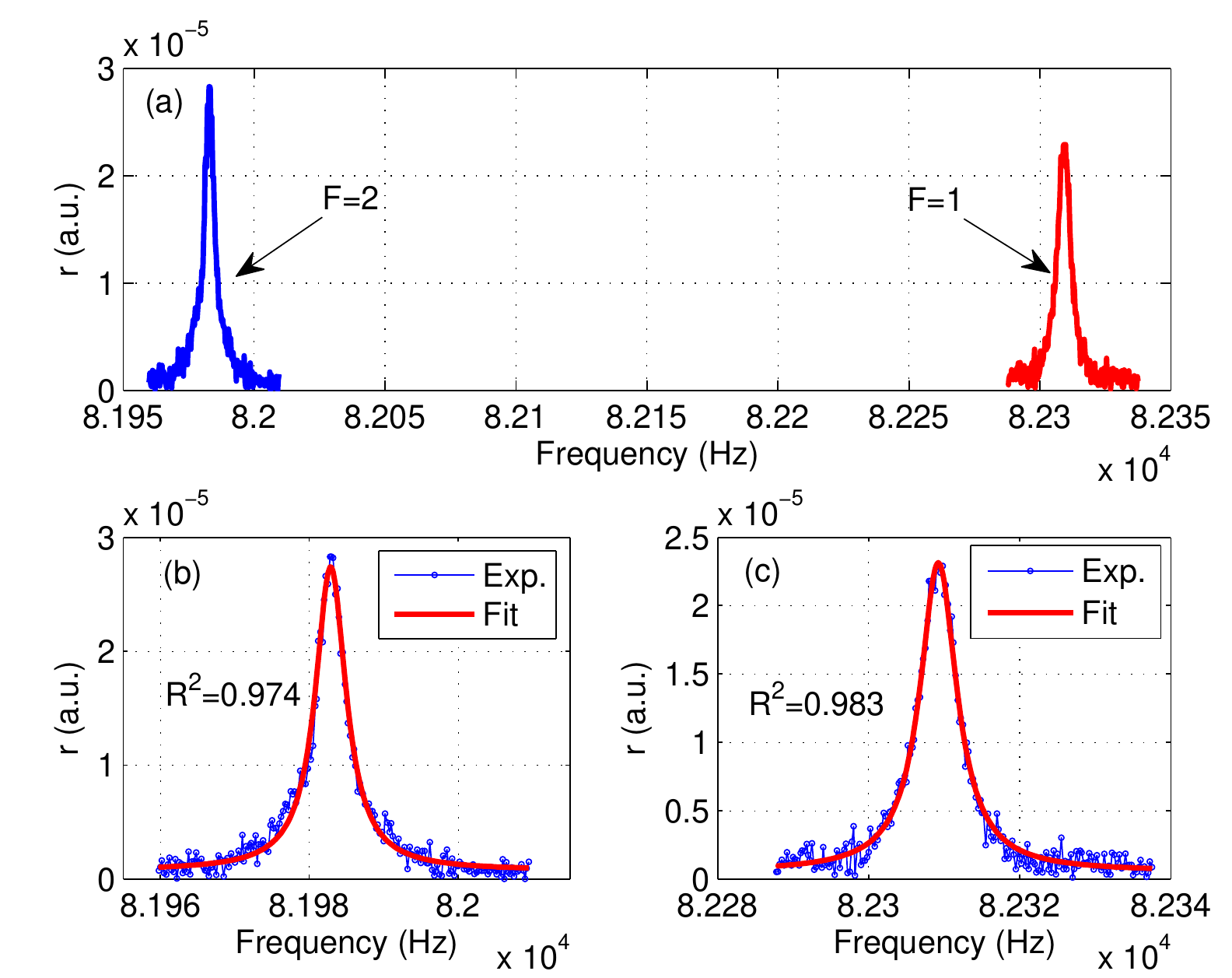}
\quad
\caption{\label{fig:ZeemanSpectro}  (color online). (a) Typical Zeeman spectrum of $^{87} \rm Rb$ atoms at $11.72 \ \rm \mu T$, acquisitied by recording r (R-channel of lock-in amplifier) when scanning frequency of $B_{\rm{x}}$. The gyromagnetic ratio of the $F=1$ hyperfine level in ground state is $\gamma_1= 7023.69\ \rm{Hz/\mu T}$ and that of $F=2$ is $\gamma_2= 6995.83\ \rm{Hz/\mu T}$. As a result, they can be distinguished clearly. (b) Fitting curve with Lorentzian profile for $F=2$ magnetic resonance signal. The legends Exp. and Fit denote experimental data and fitting curves, respectively. (c) Fitting curve for $F=1$ magnetic resonance signal. The Zeeman spectrum is a bit asymmetric, mainly because the frequency scanning is a transient process. For these data, the probe laser power is $200\mu$W.}
\end{figure}

We change the laser power by a neutral density filter and change the laser frequency by tuning the electric current flowing through the DFB laser diode, so that the influences of these laser parameters on frequency ratio are obtained. The transverse relaxation time of the polarized atoms is approximately 60 ms according to the resonance curve in Fig. 2, and it takes approximately 1 s to scan through the resonance peak, so the spins are not in good equilibrium. As a result, the scanning curve is not an ideal Lorentzian profile. Taking this issue into account, we used a group of four phases to obtain a value of frequency ratio, as shown in the Supplemental Material \cite{Zhiguo_2019}. The normalized frequency ratio (NFR), namely $\nu_2/\nu_1 -0.996033924$, as a function of laser frequency and power, is shown in Fig. 3.
The NFR is on the order of $10^{-7}$ over an optical frequency range of 7 GHz centered on the $^{87} \rm Rb$ $\rm D_2$  line.
 The dispersion of NFRs at $450 \ \rm \mu W$ is slightly larger than that at $200 \ \rm \mu W$, which may be due to the residual light shift. As a whole, however, the measured NFR shows low sensitivity to the variations of laser frequency and power.

\begin{figure}[h]
\includegraphics[width=3.3in]{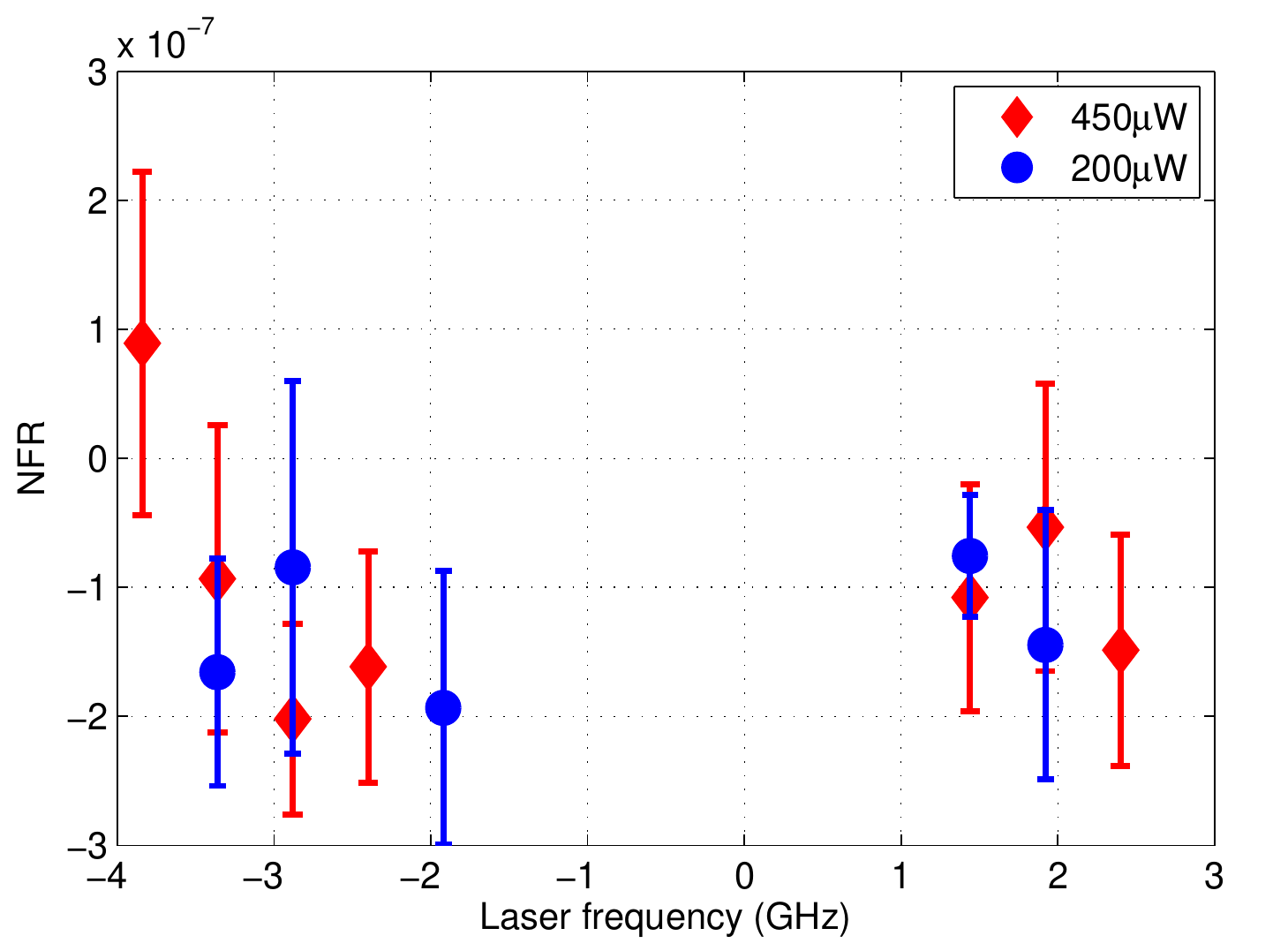}
\caption{\label{fig:RvsLaserfreq} (color online). Measured NFR ($\nu_2/\nu_1 -0.996033924$) as a function of laser frequency for laser powers of $200 \ \rm \mu W$ and $450 \ \rm \mu W$, demonstrating low sensitivity to variations of laser frequency and power. The laser frequency in the x-axis is shifted relative to the transition frequency from $5^{2}S_{1/2} (F = 1)$ of the $^{87} \rm Rb$ $\rm D_2$ line. At some frequencies, only one of the hyperfine magnetic resonance signals was larger than noise background, so the frequency ratio could not be obtained. The laser power was measured at the entrance hole of the magnetic field shield. }
\end{figure}

To check the ability of the comagnetometer to suppress the magnetic field variation, we measured the NFR at different $B_0$ values, as shown in Fig. 4. We did not find any trend of NFR upon changing the strength of the bias magnetic field, except the fluctuations on the order of $10^{-7}$ due to the statistical error. At our experimental parameters, the influence of the nonlinear Zeeman effect on NFR is much less than $1.5\times10^{-9}$ \cite{Zhiguo_2019}.

\begin{figure}[h]
\includegraphics[width=3.3in]{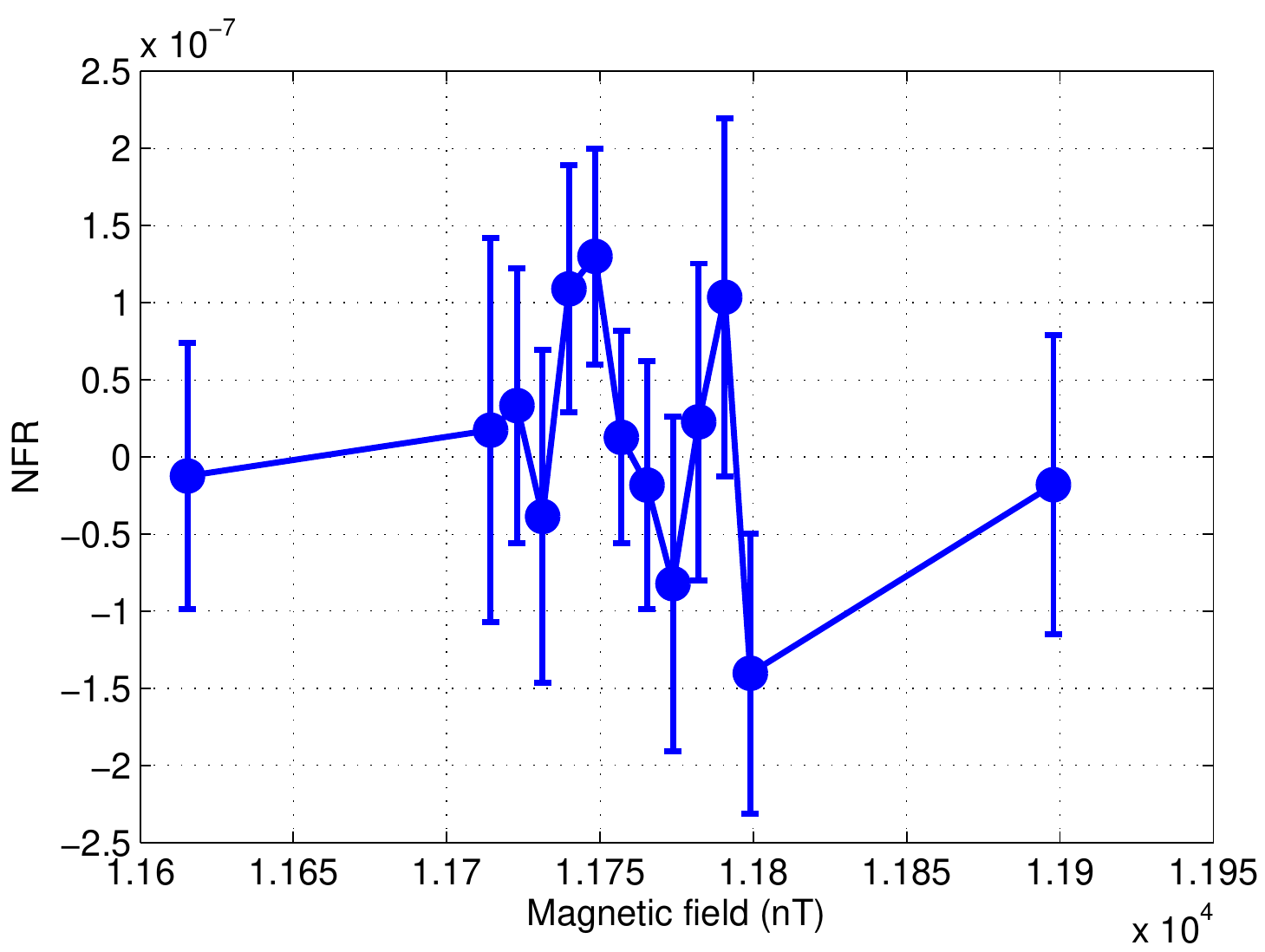}
\caption{\label{fig:RvsMF} Measured NFR as a function of bias magnetic field $B_z$, demonstrating extremely low sensitivity to magnetic field variation. $B_z$ is changed by tuning the current flowing through the $z$ coils. At first, we obtained the middle 11 points, which do not show the NFR trend with magnetic field. Therefore, we measured the NFR with a slightly larger magnetic field variation; that is, the two points at the sides, which also do not show the NFR trend with magnetic field. For these data, the probe laser power is 200 $ \mu $W. }
\end{figure}

Magnetic field gradient can lead to systematic errors in the comagnetometers, especially for those using dual-species spins \cite{PhysRevLett.113.163002, Ledbetter_2012}. For our comagnetometer, using only $^{87} \rm Rb$ atoms, the magnetic-gradient effect resulting from the spatial separation between the ``centers of spin'' is nearly eliminated. The NFR as a function of magnetic field gradient $\partial B_{\rm{z}}/\partial \rm{z}$ is shown in Fig. 5. It shows no apparent trend depending on magnetic field gradient, except statistical fluctuations on the order of $10^{-7}$.

\begin{figure}[h]
\includegraphics[width=3.3in]{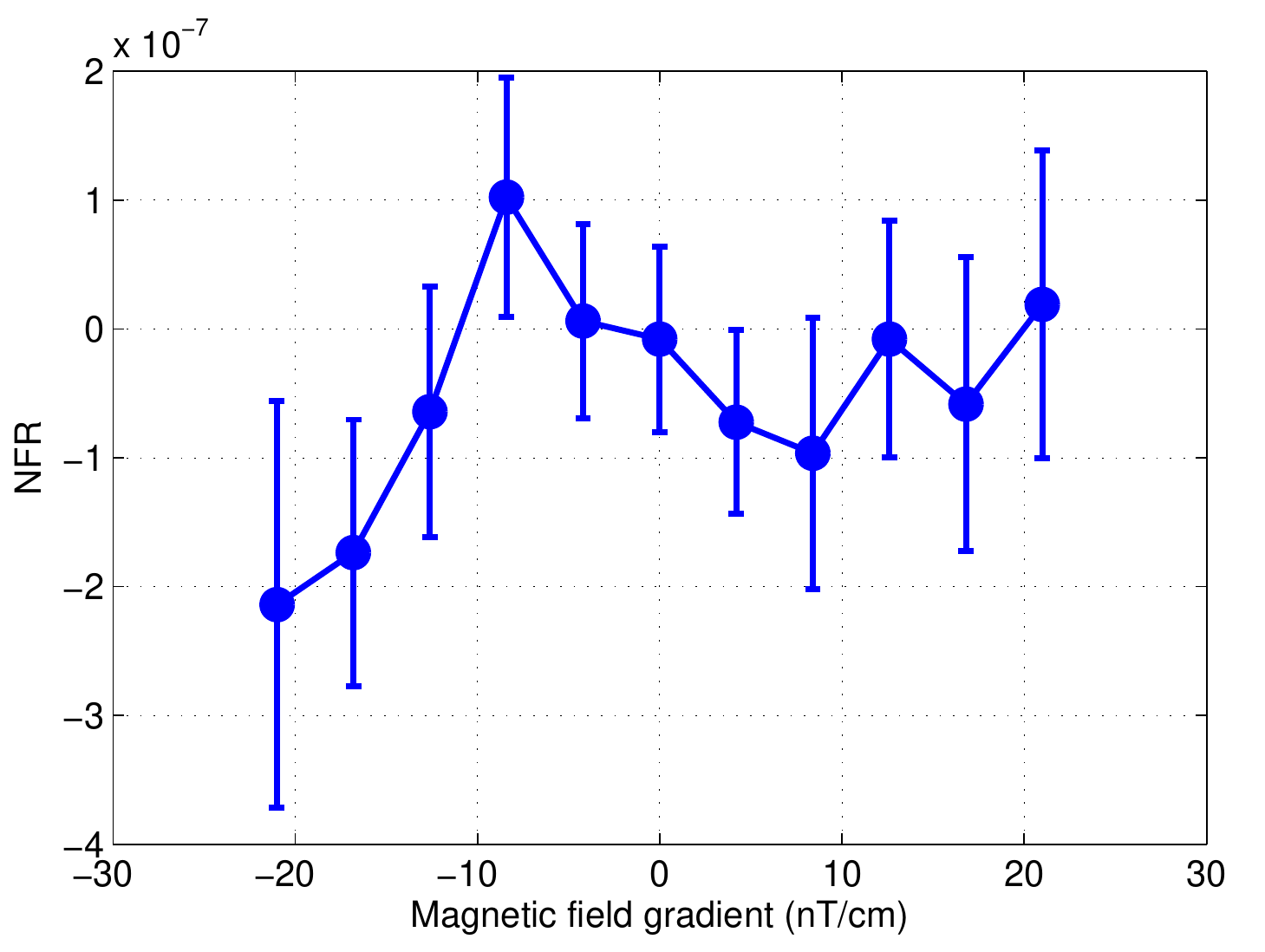}
\caption{\label{fig:epsartDaniel2015}   NFR as a function of magnetic field gradient  $\partial B_{\rm{z}}/\partial \rm{z}$ , demonstrating low sensitivity to magnetic field gradient variation. We use anti-Helmholtz coils to produce $\partial B_{\rm{z}}/\partial \rm{z}$, the magnitude of which is changed by controlling the current.}
\end{figure}

Because of the simplicity in structure and small systematic errors, our comagnetometer is a good device for tests of fundamental physics, such as spin-gravity coupling. The gyrogravitational ratios of $^{87} \rm Rb$ are given in the following equations \cite{Kimball_2015},
\begin{equation}
 \chi_{F=2}=0.25\chi_{\rm{e}}+0.25\chi_{\rm{p}}, \label{eq:1}
\end{equation}
\begin{equation}
 \chi_{F=1}=-0.25\chi_{\rm{e}}+0.42\chi_{\rm{p}}, \label{eq:2}
\end{equation}
where $\chi_{\rm{e}}$ and $\chi_{\rm{p}}$ are the gyrogravitational ratios of electrons and protons in $^{87} \rm Rb$ atoms, respectively.

The spin-precession frequencies of the two hyperfine levels, are given by
\begin{equation}
\nu_1\left(  \pm  \right)=\gamma_1 B_{\rm{z}}\mp\chi_{F=1}\frac{g\cos \phi}{\hbar}, \label{eq:3}
\end{equation}
\begin{equation}
\nu_2\left(  \pm  \right)=\gamma_2 B_{\rm{z}}\pm\chi_{F=2}\frac{g\cos \phi}{\hbar}, \label{eq:4}
\end{equation}
where ''$\pm$'' in $\nu_1\left(  \pm  \right)$ and $\nu_2\left(  \pm  \right)$ denote reversing the magnetic field direction, $g$ is the acceleration due to gravity, and $\phi$ is the angle between the bias magnetic field $B_{\rm{z}}$ and Earth's gravitational field. The opposite signs of the $\chi_{F=1}$ and $\chi_{F=2}$ terms in the right-hand sides of (\ref{eq:3}) and (\ref{eq:4}), respectively, derive from the corresponding opposite Larmor-precession directions of the $F=1$ and $F=2$ hyperfine levels. We construct the following ratios,
\begin{equation}
R_+=\frac{\nu_2\left(+\right)}{\nu_1\left(+\right)}=\frac{\gamma_2B_{\rm{z}}+\chi_{F=2}g\cos\phi/\hbar}{\gamma_1B_{\rm{z}}-\chi_{F=1}g\cos\phi/\hbar}, \label{eq:5}
\end{equation}
\begin{equation}
R_-=\frac{\nu_2\left(-\right)}{\nu_1\left(-\right)}=\frac{\gamma_2B_{\rm{z}}-\chi_{F=2}g\cos\phi/\hbar}{\gamma_1B_{\rm{z}}+\chi_{F=1}g\cos\phi/\hbar}. \label{eq:6}
\end{equation}

The difference of the ratio obtained by reversing the magnetic field $B_{\rm {z}}$ is
\begin{equation}
\Delta R=R_+-R_-\approx\frac{1.34\chi_{\rm{p}}}{\gamma_1 B_{\rm{z}}}\frac{g\cos\phi}{\hbar}. \label{eq:7}
\end{equation}

\begin{figure}[h]
\includegraphics[width=3.4in]{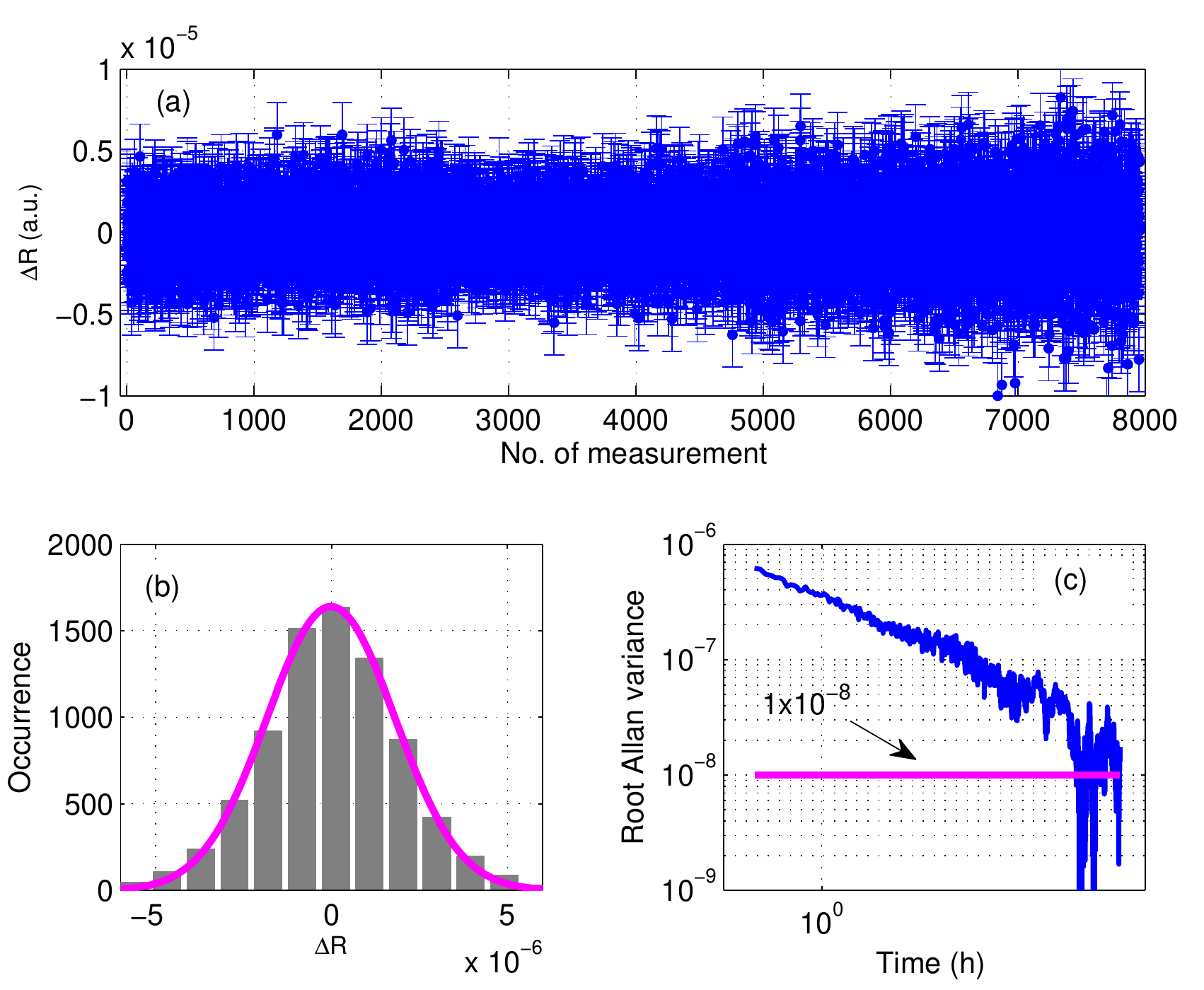}
\caption{\label{fig:Data3in1} Results of spin-gravity coupling measurement over a period of 16000 minutes. (a) The calculated $\Delta R$ based on the measured $R_{ \pm  }$. (b) Histograms of $\Delta R$ with Gaussian fitting: solid lines in the histograms are Gaussian-fitting curves, indicating $\Delta R$ is subject to normal distribution. (c) Allan variance analysis for $\Delta R$, showing a slight drift at the 100-h timescale. The measured $\Delta R$ scatters somewhat larger after 5000 points mainly due to fluctuations of the gas flowing from air-conditioner in the laboratory, which blows the experimental setup.}
\end{figure}

The experiment was performed at ChangSha, China ($28^\circ$N,$113^\circ$E). We placed the experimental setup with the x axis along a West-East direction, the y axis along a South-North direction, and the z axis was directed to the sky. To make the residual Stark shift as small as possible, we set the laser power as $50\mu$W and detune the laser frequency more than 2.5 GHz far from $F\rightarrow F'$ transitions of the $^{87}$Rb $D_2$ line by stabilizing its frequency to the $5^2P_{1/2}$ $(F=2)$ to $5^2P_{3/2}$ transition of $^{85}$Rb atoms, though it reduces signal-to-noise ratio (SNR) to some extent. We used eight phases to obtain one set of data so as to reduce the transient effect in the scanning process  \cite{Zhiguo_2019}. In the experiment, we collected 8,000 groups of data at a rate of 30 groups per hour, as shown in Fig. 6. The measurement shows $\Delta R=(6\pm22_{stat})\times 10^{-9}$ with Gaussian analysis and $\Delta R=1.0\times 10^{-8}$ with Allan variance analysis. The data above correspond to the spin-dependent gravitational energy of the proton at a level of $2(8)\times 10^{-18}$ eV and $4\times 10^{-18}$ eV, respectively.

\section{\label{sec:level3} Discussion}

The single-species comagnetometer essentially has fewer systematic errors, since its sensitivity to magnetic field gradient and variation of laser power and frequency is extremely small in theory. At present, it gives comparable sensitivity to the most stringent constraints on the proton spin-dependent gravitational energy \cite{PhysRevLett.121.023202, PhysRevD.96.075004}. However, it is still far from reaching its fundamental limit, with $\Delta R$ of the order of $10^{-10}$ for a measurement time of $1.6\times 10^{4}$ minutes \cite{Zhiguo_2019}. The main reasons for the huge gap between fundamental sensitivity and practical sensitivity are given as follows. (a) We use the scanning method to get the Larmor frequency, with which we acquire the two hyperfine magnetic resonance curves time-divided. Thus, each phase takes 15 seconds with only one second scanning through resonance, which decreases the measuring efficiency and meanwhile reduces noise rejection bandwidth and effect. As a result, the technical noises due to current source, air flow fluctuations cannot be common-mode suppressed effectively. (b) To reduce the possible light shift, we set the laser power at 50 $\mu$W. The SNR was only approximately 10 in the present experiment system, which is somewhat lower compared with other comagnetometers. To improve the practical performance of the single-species comagnetometer, a better current source for bias magnetic field \cite{Inoue_2016} and a system-level temperature control with reducing the air flow fluctuation. Besides, by optimizing the system parameters such as cell temperature, laser power, amplitude of driving magnetic field \cite{Domenico_2007}, we could get closer to the fundamental limit.

For the problems of low measuring efficiency and finite noise-rejection bandwidth in the scanning method, we tried to operate the Rb atomic spins as a dual-maser mode. The preliminary experiment has shown nearly a 2-orders-of-magnitude enhancement in sensitivity, but there was frequency bias due to pumping-light and feedback-loop phase shift \cite{Zhiguo_2019}. In this case, modulating $\Delta R$ with the position of a non-magnetic spin-dependent sample (e.g., a zirconia rod \cite{PhysRevLett.111.102001}, Pb source masses \cite{lee2019new}) instead of bias magnetic field may be more effective.

\section{\label{sec:level4} Conclusions}

We have realized a new kind of comagnetometer based on the Zeeman transitions in the hyperfine levels of $^{87}\rm Rb$ atoms. With a simple structure, the single-species magnetometer demonstrated an excellent ability to suppress magnetic field variation and the systematic error caused by the magnetic field gradient. Preliminary experimental results show that it is one of the most sensitive devices in spin-gravity coupling measurements. Through further improvements on the system, such as choosing a better current source, optimizing the system parameters, and making the $^{87}\rm Rb$ atoms be continuous masers, the sensitivity could be significantly improved.

\begin{acknowledgments}
This research was supported by the Natural Science Foundation of China (Grant Nos. 61671458, 61571018, 61571003, and 91436210), the Natural Science Foundation of Hunan Province (Grant No. 2018JJ3608), and the Research Project of National University of Defense Technology (Grant No. ZK17-02-04). One of the authors (Z.W.) thanks Dr. Teng Wu for helpful discussions.
\end{acknowledgments}

\bibliography{comagnetometer}

\begin{thebibliography}{44}
\bibitem [{Zhi()}]{Zhiguo_2019}%
  \BibitemOpen
  \href@noop {} {\bibinfo {title} {See {S}upplemental {M}aterial at [url] for
  the detailed discussions of possible error sources: the higher-order {Z}eeman
  effect, which includes {R}ef. [22]; the {Z}eeman light shift, which includes
  {R}efs. [18, 23-28, 40]; the experiment, which includes {R}efs. [1, 29]; the
  fundamental limit, which includes {R}efs. [30-32, 42-43]; the dual-maser
  operation, which includes {R}efs. [10, 33-34]; the magnetic gradients, which
  includes {R}efs. [13-14, 35-36].}}\BibitemShut {Stop}%
\end{thebibliography}

\end{document}